# Purely Mechanical Memristors: Perfect Massless Memory Resistors, the Missing Perfect Mass-Involving Memristor, and Massive Memristive Systems

Sascha Vongehr[a]

National Laboratory of Solid State Microstructures, College of Modern Engineering and Applied Sciences, Nanjing University, PRC

We define a mechanical analog to the electrical basic circuit element $M = d\varphi/dQ$, namely the ideal mechanical memristance $M = dp/dx$; $p$ is momentum. We then introduce a mechanical memory resistor which has $M(x)$ independent of velocity $v$, so it is a perfect (= not-just-memristive) "memristor," although its memristance does not crucially involve inert mass. It is practically realizable with a 1cm radius hollow sphere in heavy fuel oil with a temperature gradient. It has a pinched hysteretic loop that collapses at high frequency in the $v$ versus $p$ plot. The mechanical system clarifies the nature of memristor devices that can be hypothesized on grounds of physical symmetries. We hypothesize a missing mechanical perfect memristor, which must be crucially mass-involving (MI) precisely like the 1971 implied EM memristor device needs magnetism. We also construct MI memristive nano systems, which clarifies why perfect MI memristors and EM memristors are still missing and likely impossible.

## 1   Introduction: The Significance of Mechanical Memristors

 The 1971 predicted[1] perfect real memristor device is a significant hypothesized scientific entity in need of verification or disproof much like Dirac's positron in 1928 or magnetic monopoles. It is based on fundamental symmetries that are on the same profound level as Murray Gell-Mann's SU(3) group-symmetry representations predicting subatomic particles, partially because magnetism is a relativistic effect instead of an independent charge.[2,3] The 2008 claimed[4] and widely publicized discovery was rigorously rejected on those fundamental theoretical grounds,[2,3] but the arguments seem to be prohibitively unusual in the engineering oriented research community. They have been very much clarified with the first construction of purely mechanical memristors[3] where position $x$ is the charge so that velocity $v = dx/dt$ is the current. The mechanical

---

[a] Corresponding author's electronic mail: vongehr8@yahoo.com

circuit theories with and without inert mass *m* are *by construction* mathematically precise analogs of the electrical circuit theories with and without induced magnetism. However, the mechanical analog has many advantages:

**1)** While magnetic monopoles and entirely different magnetisms seem too speculative, mass is "switched on" in different ways by adding inertia and/or gravity in the system. **2)** Historically, inductor L was a known third kind of device that led to a hypothesized fourth, the missing M, and this then led to the claimed discovery, let us call it "claimed-M". The massless mechanical perfect memory resistor is the exact mechanical analog of "claimed-M," yet it is naturally found *before* introducing inert mass, which is the analog of the real inductor device L. The historical sequence is turned upside down in the analog, but without such being any longer a mere speculative thought experiment as still in 2012.[2] If missing-M and claimed-M were the same, this M would lead to us missing a device L precisely like historically L led to our missing M (on grounds of the same symmetry). The mechanical analog shows that the statement "claimed-M = missing-M" is equivalent to claiming inert mass without inert mass. The hypothesized, missing mechanical perfect memristor must self-evidently be crucially mass-involving (MI), precisely like the 1971 implied memristor device requires magnetism (thus "EM memristor"). **3)** MI memristive systems clarify why perfect MI or EM memristors are still missing and likely impossible. They thus also indicate where the missing memristors could perhaps be found if they are possible after all! This all makes purely mechanical memristors highly interesting and widely relevant before even considering possible applications of such devices and theories in nanotechnology. Hence, this paper focuses on mechanical memristors. We re-derive them pedagogically, show that the correct hysteretic behavior is obtained from a practically easily realizable system. We then introduce a new MI memristive system that makes *m* dependent on $x^\alpha$ with the power α being adjustable as desired, which may help to either strictly show that the missing memristor devices are impossible or to finally discover a missing memristor.

## 2  Terminology

"Memristor" refers in the literature variously to memory resistor, memristive system, perfect memristor, the theoretical 'basic two-terminal circuit element' (BCE), and others.

Given the widespread confusion, we must commit to one strict terminology. Given the wide audience that memristors attract, "ideal" must be the opposite of "real" for example; it should not sometimes mean "perfect." The 'ideal resistor' $R$, 'ideal capacitor' $C$, 'ideal inductor' $L$, and 'ideal memristor' $M$ (also 'resistance,' 'capacitance,' 'inductance,' and 'memristance') are the four BCE of electrical circuit theory. BCE are *basic* because they are *independent* of each other like a *basis* of four linearly independent vectors. One cannot connect ideal resistors together but end up with capacitance. It is therefore that real devices are never ideal. Two metal plates make a *real* (non-ideal) capacitor *device*; but it has always some resistance, too. Moreover, the BCE are *passive* and were called "passive circuit elements,"[5] meaning they do not supply energy. Violating passivity will violate independence. The BCE are *theoretical ideal* entities and strictly speaking all impossible as real devices. Their relevance rests in theoretical modeling and they exist by definition; they do not need to be discovered! In 1971, a real memristor *device* was hypothesized.[1] It is called the *missing fourth,*[6] because real resistor, capacitor, and inductor devices do exist. The 'perfect memristor'-versus-'memristive system' distinction was defined only in 1976.[7] Charge controlled systems $M_{(Q)}$ have been called ideal/real/true/genuine/perfect memristors, because $M$ depends not also on current $I$ for example. We stick to calling them 'perfect memristor.'[4] A memristive (non-perfect) system can depend also on the current $I$ for example, thus $M_{(Q,I)}$.

## 3 Circuit Theories without Magnetism and Mass

The *current-charge relation* $I = dQ/dt$ is the *first fundamental relation* (FR1) of circuit theory. It defines current $I$ as the time derivative of charge $Q$. It can be defined for mass flows for example. In our mechanical analog, the "charge" is position $x$, which flows past a moving body and is conserved behind it ('conserved charge'). FR1 is velocity $v = dx/dt$. Electrical charges have force fields between them. Hence, charge storage will store a corresponding energy which can also be dissipated in the charge flow. In the electrical case, this is described by voltages $U$. Charge storage leads to the first BCE, $C = dQ/dU_C$. The mechanical analog of a real capacitor device is to "store" $x$ in the displacement of a spring with Hooke's spring stiffness $k$. The BCE is therefore $1/k = dx/dF_k$. Dissipation is modeled by $R = dU_R/dI$. Our mechanical resistor device is a light hollow sphere submerged in oil in an orbiting satellite (no gravity, no buoyancy). There are large composite mechanical circuits such as the Rouse model in polymer dynamics, a chain of springs and beads in a viscous fluid, which is the analog of Lord Kelvin's discrete LC chain model of the transatlantic telegraph cable. The mechanical and electrical circuit theories are mathematically precise analogs by construction; no further similarity between mechanics and electrics is claimed or necessary.

With sufficiently viscous oil and low speeds, the oil flow around the sphere is laminar and the friction force $F_f$ proportional to $v$. The drag coefficient $c = dF_f/dv$ is the BCE. The circuit couples these forces, for example if the body in oil is dragged by the spring. A triangular symmetry, namely the red base-triangle of the tetrahedron in Fig. 1a, thus connects the three fundamental circuit variables $Q$, $I$, and $U$ ($x$, $v$, and $F$, respectively), and the BCE are more generally written as $C_{(U,Q)} = dQ/dU$, $R_{(U,I)} = dU/dI$, etc.

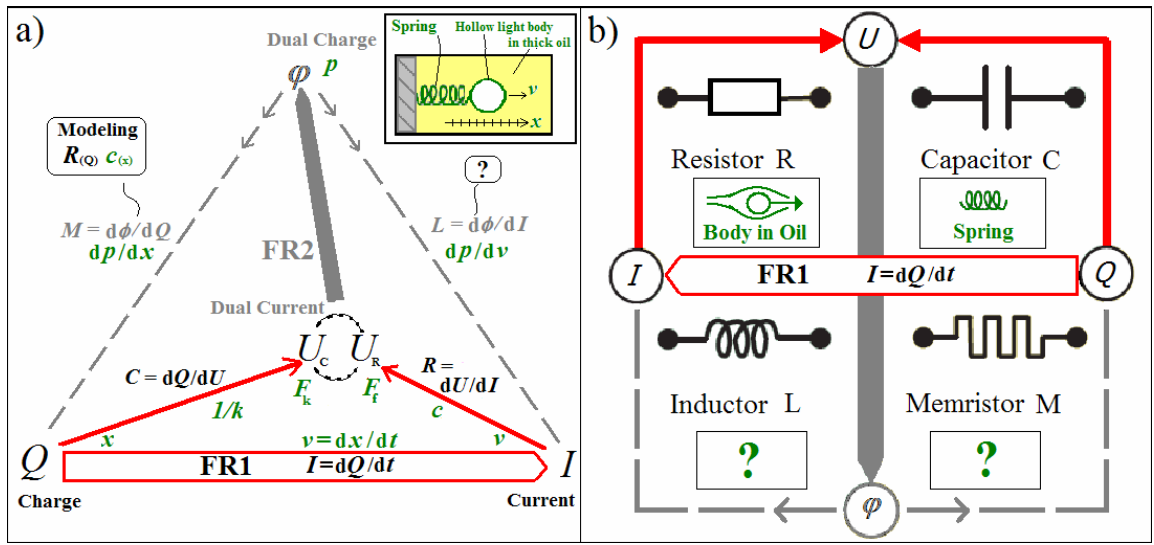

**Figure 1:** The symmetry of the fundamental circuit variables; (a) the chains indicate that the voltages $U_C$ and $U_R$ are coupled by the circuit; the mechanical analog of a light hollow sphere in oil is inset; its variables are shown in green. The mass of sphere and spring are negligible, so the system is strongly over-damped and cannot oscillate. FR2 with the fourth variable $\varphi$ ($p$ in the mechanical system) erect a tetrahedron on top of the red $Q$-$I$-$U$ base-triangle. As do $R$ and $C$ above FR1, $L$ and $M$ label two new edges which correspond to two new BCE. The mechanical system's $L = dp/dv$ has units of kg, but the system's sphere does not have an inert mass yet (one would use different units when growing up in that universe). (b) $M$ and $L$ are on the same footing and *both* still equally absent; without magnetism (or mass in the mechanical system) there are no corresponding devices yet. The "inductor" cannot be the EM inductor device. $M$ and $L$ help modeling but suggest no new devices. The arrows indicate that physical charge is prior to the definition of current, and the general force terms $U$ and $F$ are discovered via the devices.

### 3.1 Absence of magnetism

Maxwell's equations (ME) come in two pairs. The first pair (MEP1) relates the free charge and current densities: $\{\nabla\cdot\vec{D} = \rho_{\text{free}}; \nabla\times\vec{H} = \vec{j}_{\text{free}} + d\vec{D}/dt\}$. The second pair is the first pair's magnetic twin, but there are no magnetic charges because magnetic fields are a relativistic correction: $\{\nabla\cdot\vec{B}=0; \nabla\times\vec{E} = -d\vec{B}/dt\}$ (MEP2). Magnetism is due to Lorentz-Fitzgerald contraction and time dilatation on moving charge distributions. The second equation of MEP1 can be written $\nabla\times\vec{B} = \mu_0\vec{j}_{\text{free}} + \varepsilon_0\mu_0 d\vec{E}/dt$; and $c_0 = (\varepsilon_0\mu_0)^{-1/2}$ is the velocity of light. If $\mu_0$ were much smaller, we would not know magnetism. In the mechanical analog, the thick oil renders the mass of the body unnoticed. Neglecting magnetism is equivalent to concentrating on the non-relativistic limit $c_0 \to \infty$ as done in much of mechanics, classical and quantum. Neglecting magnetism is what the concept of independent BCE is partially about! There are always magnetic fields with any current, but circuit theory models RC-circuits usually without mentioning *L*. Differently put, especially circuit theory finds neglecting magnetism unproblematic. For now, with *B* and *H* negligible, only parts of MEP1 remain. Integration of $\rho_{\text{free}}$ and $j_{\text{free}}$ results in charge *Q* and current *I* and thus FR1. In other words, also non-magnetic electrical circuit theory does derive from Maxwell theory, but MEP2 is not involved. The circuit theory discussed here does not know about magnetism.

### 3.2 Flux, *L*, and *M* without magnetism and a mechanical massless "memristor"

Defining a "flux" $\varphi = \int U dt$ provides a so called second fundamental relation, $U = d\varphi/dt$ (FR2). Unlike the very similar equation further below, it does *not* derive from MEP2. It is defined this way because *φ* is thereby a canonically dual charge, because the force term *U*, which relates to energy, is the dual current (compare FR2 to FR1). Current is a time derivative, and energy *E* and its canonical dual time *t* are the main players in dynamical physical theories. Circuit theory rests on this because energy conservation and charge conservation lead to Kirchhoff's loop and node rule, respectively. The mechanical equivalent is therefore the canonical dual to *x*, namely momentum $p = \int F \mathrm{d}t$ leading to $F = \mathrm{d}p/\mathrm{d}t$ as the FR2. Apart from the new $U$-to-$\varphi$ edge, there are thus again two more edges, almost as if we introduced another charge with a force field: $\varphi$-to-$I$ and $\varphi$-to-$Q$. These correspond to two further binary relations, and thus two more BCE can be defined: $L_{(\varphi,I)} = \mathrm{d}\varphi/\mathrm{d}I$, and $M_{(\varphi,Q)} = \mathrm{d}\varphi/\mathrm{d}Q$ relates the dual charges. Such a tetrahedral construct provides complete circuit theory in the sense of that the four BCE can model all potentially non-linear circuit behaviors of the fundamental circuit variables.

Because of $\mathrm{d}\varphi/\mathrm{d}Q = (\mathrm{d}\varphi/\mathrm{d}t)/(\mathrm{d}Q/\mathrm{d}t) = \mathrm{d}U/\mathrm{d}I$, memristance is a resistance with standard units of Ohm, $[R] = \Omega$, and a linear memristor $M = \varphi/Q$ is a constant Ohmic resistor. Independence between BCE therefore requires $M$ to be non-linear, as all the BCE can generally be. The mechanical analog is now obvious, and we hereby describe a new, mechanical ideal memristor BCE with $M$ being a non-linear $\mathrm{d}p/\mathrm{d}x$ having the units of drag resistance, $[c] = \mathrm{kg/s}$. $M$ is a resistance that depends on $Q_{(t)}$; it memorizes the charge that has flown through it; hence "memristance." It facilitates modeling charge dependent resistors $R_{(Q)}$. The mechanical analog of this is $c_{(x)}$, for example if the oil's dynamic viscosity $\eta$ depends on $x$. Viscosity is modeled with the Andrade equation $\eta = A\, e^{B/T}$.[8,9] A heavy fuel oil (HFO) such as HFO-380 is for our purposes sufficiently well modeled with $A = 0.2\ \mathrm{kg\ m^{-1}\ s^{-1}}$ and $B = 47°\mathrm{C}$. If the $x = 0$ end of the oil bath is held at 20°C while $x_{\max} = 1\ \mathrm{m}$ is held at 30°C, the gradient is approximately $T = 20°\mathrm{C} + (10°\mathrm{C/m})x$. The resulting $\eta_{(x)}$ is plotted in Fig. 2.

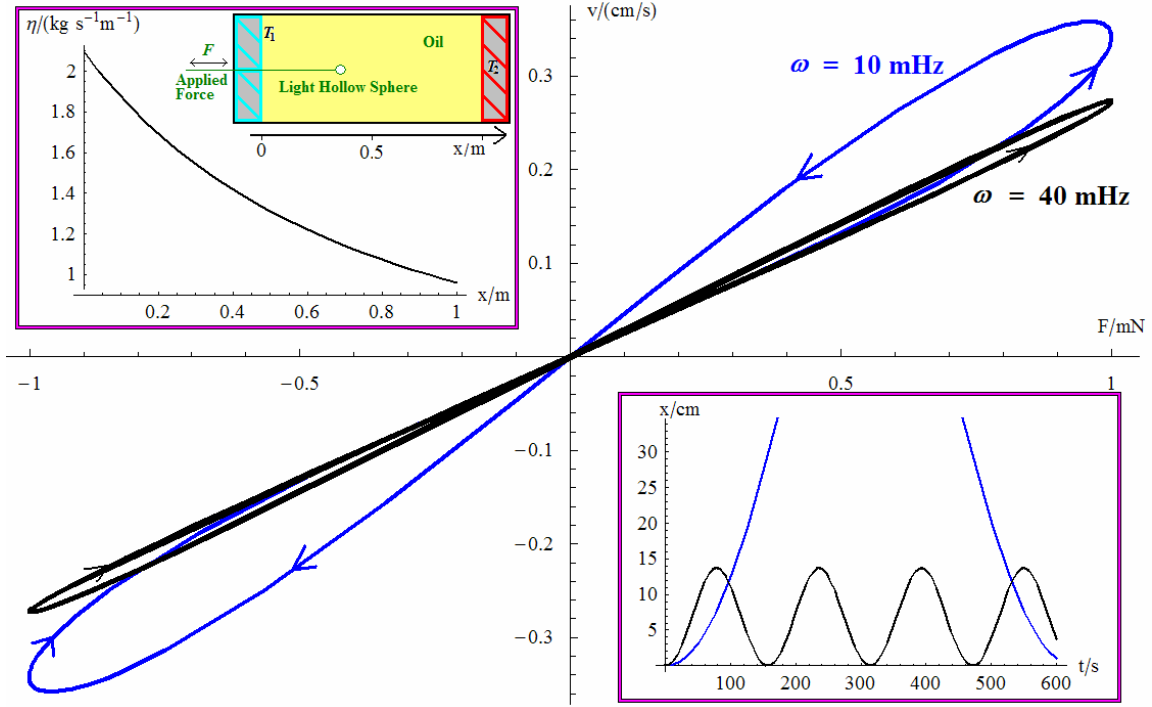

**Figure 2:** Simulation of the purely mechanical perfect memristor with the numerical NDSolve function of Mathematica5®. During the ten simulated minutes, the blue loop is just about almost completed while the black loop is almost passed through four times. The lower inset shows position *x* versus time; the blue curve peaks at 70 cm. The upper inset illustrates the system schematically and shows the oil's viscosity $\eta$ as it depends on *x*.

The body is spherical, thus $c = 6\pi\eta r$. The radius is taken to be $r = 1$ cm. If applied forces stay below $F_{max} = 1$ mN, speed $v_{max} = F_{max}/c_{(1m)}$ cannot exceed 5 mm/s. The oil's density $\rho = 0.9$ g cm$^{-3}$ changes comparatively little with *T*. The Reynolds number $\text{Re} = \rho v(2r)/\eta$ is thus below 0.1 and the oil flow always laminar. Many researchers call any system "memristor" if it has a pinched hysteretic loop in the *I*-vs-*U* plot. The mechanical system shows this pinched hysteretic loop in the corresponding *v*-vs-*F* plot if the sphere starts at the rest position at $x = 0$ and a force $F_{max}\sin(\omega t)$ with frequency $\omega = 10$ mHz is applied (Fig. 2). At 40 mHz, the loop is much narrower. It becomes a linear resistor at high frequency, like memristors should. Some may want to call it a force-controlled mechanical memristive system analogous to voltage controlled memristive systems[10] (compare Figure 2 of that reference or this one[4]). If all "*Resistance switching memories are memristors,*"[11] it is a "memristor." The $c_{(x)}$ does not need to be written as $c_{(x,v)}$, so this is a *perfect* "memristor," the precise equivalent to the device that was in 2008 claimed to be the missing memristor!

### 3.3 No missing real devices are suggested without magnetism or mass

We described not just electrical circuit theory without (noticeable) magnetism but also an equivalent mechanical circuit theory without (noticeable) mass. Therefore, *L* does not yet correspond to any device! *L* will become something physical below, but whether and how such exists depends on the details of the system, the universe we (or the hollow sphere) are in. The grounds on which the real memristor device was proposed is still absent. A real EM inductor device (or crucially mass-involving (MI) inductor device) *cannot* be known yet, but this *third* device is vital to the originally predicted *missing fourth*.[6] The 1971 proposal cited Mendeleev's 1870 prediction as a relevant precedent, because that hypothesis rested on empty cells in the periodic table of the elements[12] just like the missing memristor device was a vacant cell in the table Fig. 1b. In hindsight, some may object that our purely mechanical perfect "memristor" is a third device and that a fourth kind of device can now already be hypothesized. However, one would *not* regard such "memristors" as a new kind of device, because they are no more than nonlinear, charge dependent resistors. Moreover, *L* and *C* are opposing edges in the tetrahedron just like *M* and *R*. Hence, an inductor is in a sense a dual-capacitor, storing dual-charge. Nevertheless, inductors are clearly not just nonlinear, somehow merely mathematically dual-charge dependent capacitors! The memristor can only be hypothesized as an interesting (potentially different kind of) device if the three known devices, namely resistor, capacitor, and *EM* (or *MI*) inductor are indeed different, independent from each other.

Mass is fundamentally inertia like magnetism is EM inductance. These are clearly new kinds of phenomena. For example magnetism involves an *induced* voltage, a force that would not be without relativistic effects. Moreover, these phenomena could be conceivably different. Our magnetism, that relativistic effect we happen to observe instead of magnetic monopoles, turns out to supply an EM-dual magnetic charge that behaves like the canonically dual charge. This is a coincidence as far as circuit theory knows.

## 4 With Magnetism/Mass: Real Inductor Devices suggest a 4th Kind

In order to relate to the EM inductors that are *a third kind* of real device, flux must derive from the 'magnetic pair' of Maxwell theory. The electric field of MEP2 integrates to the magnetically *induced* voltage $U_{in}$. Integrating magnetic field $B$ results in the *magnetic* flux $\varphi_m = -\int_{Area} d\vec{f} \cdot \vec{B}$. This "flux-*linkage*" links the magnetic field to the induced voltage. The resulting *voltage-flux* relation relates $U_{in}$ to the time derivative of the magnetic flux: $U_{in} = d\varphi_m / dt$. The magnetic field adds $\varphi_m$ as a fourth corner that erects a tetrahedron on top of the triangle (Fig. 3a), much like introducing electric forces erected the triangle.

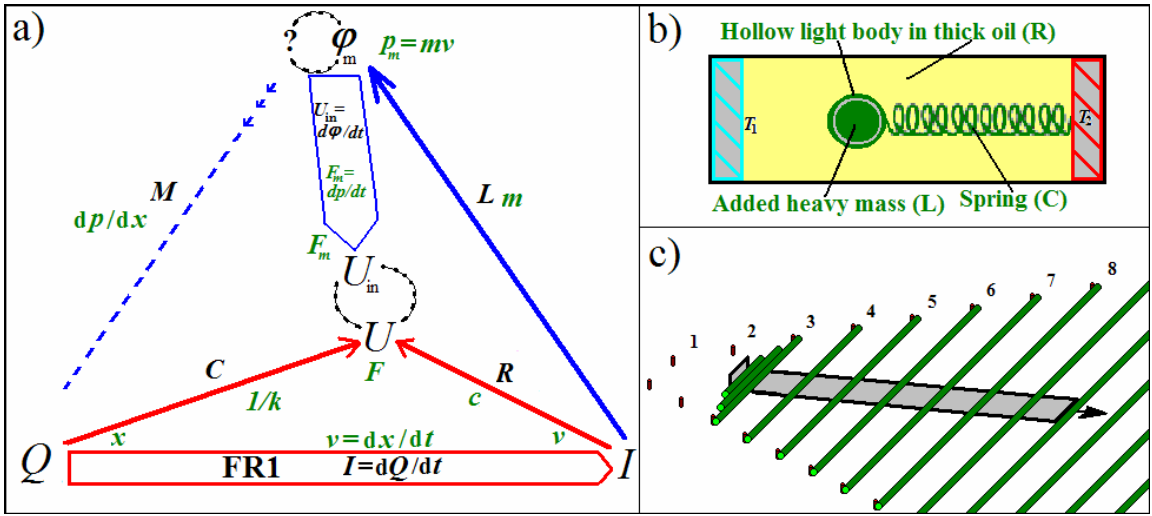

**Figure 3:** (a) The tetrahedral symmetry that suggests real memristor devices. (b) The sphere has now a heavy mass *m*, and the spring-system can oscillate like LRC-circuits. (c) A system picks up massive rods (green), depicted when adding the third rod, taking it away from its holding pins (red). d*m*/d*x* is proportional to $x^\alpha$ with suitable lengths and weights of rods. The rods stick sufficiently to add inertia also on the way back, but the sudden impact with their holding pins removes them again from the pile.

$\varphi_m$ is an EM-dual, magnetic charge, as is illustrated by calculating $\varphi_m$ around a Dirac monopole $\vec{B} \propto \vec{r}/(r^2|\vec{r}|)$ and comparing the result $B = \varphi_m/(4\pi r^2)$ with the electron's $E = e/(4\pi r^2)$. However, it is the known device, here the EM inductor, which led us to construct the tetrahedral symmetry in the first place (also historically). The symmetry *together with the real inductor* may now suggest that *M* perhaps also corresponds to a real device. However, this hypothesized device, suggested on grounds of the EM inductor, is an *EM* memristor in the sense that it must also be absent without magnetism. It must be absent without magnetism, because its existence would otherwise suggest the inductor device (precisely in the same way as the memristor was suggested to be the missing fourth next to the known third device, the inductor). Without magnetism however, a suggested inductor device cannot be the EM inductor, even if something somehow similarly behaving is found. Moreover, the circuits couple the devices; $U_C$ gives rise to $U_R$ falling along a de-charging resistor. $Q$ is made from the same charges (electrons) on the capacitor as in the resistor while de-charging. This suggests a more direct coupling between the hypothesized memristor's flux and the inductor's flux as indicated in Fig. 3a. It suggests that they should be fundamentally the same magnetic flux. The original 1971 hypothesis demanded an EM memristor: "… *the physical mechanism characterizing a memristor device must come from the instantaneous (memoryless) interaction between the first-order electric field and the first-order magnetic field…*"[1] This has become controversial, although it was precisely this which made the hypothesized memristor interesting to many. The implied EM memristor promised to be a new, *fourth kind* of device that moreover corresponds to the EM inductor like a sort of complementary EM symmetry counterpart, much like electrons suggest magnetic monopoles.

In the mechanical analog, we can switch on inertia, or if the sphere's mass $m$ was merely not noticeable in the thick oil, add a heavy mass $m$ (Fig. 3b), which is equivalent to an LRC-circuit's $L$.[5] $\varphi_m = L I$ is here mass momentum $p_m = m v$. The new force $dp_m / dt = m a = F_m$ derives from the new momentum like $U_{in}$ from $\varphi_m$. A real "dual-charge capacitor" device corresponds to $m$, namely *inert* mass. 93.9g of iridium fit into the sphere to keep it noticeably moving after suddenly switching off $F_{max}$. Given that three different kinds of real devices now correspond to all three out of four mechanical BCE, a missing fourth was hypothesized[3] precisely as done in 1971. It must be a MI memristor that cannot exist without (observable) inert mass. The position dependence of $c$ due to $T$-gradients is insufficient, because that mechanical memory resistor is already known (I); it does not require inert mass (II); it was not postulated as a fourth missing device (III); and its $dp = 6\pi r \eta_{(x)} dx$ is not inertia carrying mass momentum (IV). Its $dp$ couples to $p_m$ via the circuit (forces $F$), for instance if we attach a heavy mass at the free, left end of the forcing-lever in Fig. 2. This coupling is even more direct if $m$ is put inside the hollow sphere. However, such simply adds a MI inductor into our memory resistor. The mass is not crucially involved in the memristance. The added mass merely turns our perfect "memristor" into a memristive system.

### 4.1 Can EM and MI memristors be discovered?

Especially in the EM case there cannot be a new set of devices as if a new independent field was introduced. Magnetic monopoles would allow magnetic capacitor devices and the symmetries would be richer, but magnetism is only a relativistic effect of electrical charges. It is unsurprising that only half of the naively expected two new devices exist, namely EM inductors but not EM memristors. Momentum $p_m$ is a new charge with its force $F_m = m\,a$, not a special relativistic effect. Contraptions that pick up mass along $x$ (Fig. 3c) yield MI memristive systems, because $m_{(x)}$ makes mass crucial, and the memristance is effectively a $c_{(x,v)}$ as well as nonlinear, e.g. $(dp/dx)|_v = v(dm/dx) \propto v x^\alpha$; $\alpha \neq 0$. However, for a perfect $c_{(x)}$, mass must depend on $v$. This is generally possible, for example in special relativity theory. However, the mass must depend in just the right way and yet still be inert mass with momentum even at zero velocity although there is no mass momentum without velocity. EM memristors may conceivably have $\varphi_m$ also at zero current if magnetic fields are sustained as EM fields via electro-optics. MI memristors may need nanotechnology like our rod-gathering system in Fig. 3c, but EM memristors are rightly expected in optics or similar; so "*those interested in memristive devices were searching in the wrong places*"[4] misunderstands.

## 5  Concluding remarks

The mechanical analog illuminates a core problem: *L* and *M* are on the same, symmetrical footing in circuit theory, but the discovered "memristors," including our own perfect mechanical one, cannot deliver the grounds of hypothesizing inductors, neither EM/MI inductors *nor lesser ones*, as was explained. Disagreeing with this either implicitly claims that after finding perfect memristors, and even without ever finding any inductors, mere circuit theory predicts inert mass and relativistic magnetism. Or otherwise, and this is the apparent consensus today, one claims that the whole issue was never more than mere circuit theory, and that magnetism and mass are merely interesting ways of realizing inductors. This is clearly not true, neither historically, nor do we have rigorous arguments for why the ratio $M = dQ_{Dual}/dQ$ does not allow an EM memristor, an interesting *fourth kind* of electro-optical device that is not just a complicated resistor. Science must keep searching or disproving. The widespread opinion about that the missing memristor has been discovered is detrimental toward that endeavor. We hope that this work, conceivably via first constructing the missing MI memristor with help of our new MI memristive rod-gathering system or similar, will help archiving the missing EM memristor or proving its impossibility.